\documentclass[
superscriptaddress,
preprint,
 amsmath,amssymb,
 aps,
prb,
]{revtex4-2}

\usepackage{graphicx}% Include figure files
\usepackage{dcolumn}% Align table columns on decimal point
\usepackage{bm}% bold math
\usepackage{mhchem}
\usepackage{xcolor}
\usepackage{soul}
\usepackage{ragged2e}
\usepackage[T1]{fontenc}
\usepackage{lmodern}
\usepackage{xr}
\usepackage[utf8]{inputenc}

\begin{document}

\preprint{}

\title{Spatiotemporal Visualization of Long-Range Anisotropic Plasmon Polaritons in Hyperbolic MoOCl$_2$}

\author{Atreyie Ghosh}
\thanks{These two authors contributed equally.}
\affiliation
{James Franck Institute, The University of Chicago, Chicago, IL 60637, USA}

\author{Calvin Raab}
\thanks{These two authors contributed equally.}
\affiliation
{James Franck Institute, The University of Chicago, Chicago, IL 60637, USA}
\affiliation
{Department of Chemistry, The University of Chicago, Chicago, IL 60637, USA}

\author{Joseph L. Spellberg}
\affiliation
{James Franck Institute, The University of Chicago, Chicago, IL 60637, USA}
\affiliation
{Department of Chemistry, The University of Chicago, Chicago, IL 60637, USA}

\author{Aishani Mohan}
\affiliation
{James Franck Institute, The University of Chicago, Chicago, IL 60637, USA}
\affiliation
{Department of Chemistry, The University of Chicago, Chicago, IL 60637, USA}

\author{Muneeza Munawar}
\affiliation
{Department of Physics, The University of Chicago, Chicago, IL 60637, USA}

\author{Janek Rieger}
\affiliation
{James Franck Institute, The University of Chicago, Chicago, IL 60637, USA}

\author{Sarah B. King}
\email{sbking@uchicago.edu}
\affiliation
{James Franck Institute, The University of Chicago, Chicago, IL 60637, USA}
\affiliation
{Department of Chemistry, The University of Chicago, Chicago, IL 60637, USA}

\date{\today}% It is always \today, today,

\begin{abstract}
 {Manipulating light at the nanoscale with minimal loss remains a central challenge for nanophotonic technologies that can be tackled by using the direction-dependent polariton modes supported by anisotropic materials. Although best known for their highly confined polaritons, hyperbolic materials can also host long-range directional polaritons, whose direct observation has remained challenging as it requires experimental techniques that combine nanometre and femtosecond spatial and temporal resolution, respectively. Here, we use time-resolved photoemission electron microscopy for direct nanoscale visualization of long-range anisotropic plasmon polariton (LRAPP) dynamics on a flake of the van der Waals hyperbolic material molybdenum oxydichloride. We directly image plasmon polaritons with propagation lengths larger than 10 µm, exhibiting an approximately three times longer propagation length and intrinsically lower optical loss than short-range polaritons previously reported on the same material. By tracking the spatiotemporal evolution of LRAPPs, we determine their phase and group velocities at the nanoscale and directly observe their reflections at flake edges. These results establish molybdenum oxydichloride as a versatile platform for integrated nanophotonics, supporting both low-loss directional transport and deeply subwavelength field confinement within a single natural material, in the visible spectral range.}
\end{abstract}

\maketitle
The development of faster, more efficient, photonic devices has driven the search for materials that can manipulate light in ways that are impossible in conventional optics. Nanophotonic technologies require materials capable of confining light on the nanoscale combined with high-speed propagation and low loss.  {Anisotropic polaritonic materials have emerged as a promising solution, as their direction-dependent dielectric response enables emergent classes of guided modes with properties unattainable in isotropic media} \cite{ma2018in-plane-ec7,Pogna2024,Chen2023}.  {In isotropic noble metal thin films, coupling of interfacial modes results in guided long-range surface plasmon polaritons, which are characterized by long  propagation lengths and are commonly localized near the metal-vacuum interface }\cite{Berini_2009,Großmann2021, Burke_Stegeman_Tamir_1985}.  {However, these polaritons lack directional control, a critical limitation for advanced nanophotonic applications. Realizing long-range anisotropic plasmon polaritons (LRAPPs), which propagate over long distances while retaining strong anisotropy in their dispersion and propagation, could enable directionally selective, low-loss plasmonic transport and distinct opportunities for on-chip routing, anisotropic sensing, and coherent light-matter interactions} \cite{doi:10.1126/sciadv.abf2690, Suriyage2025}. 

Hyperbolic materials\cite{dai2014tunable-616, ma2021ghost-2bb, taboada-gutirrez2020broad-f41, ma2018in-plane-ec7, Guo2023hyperbolic-eih},  {where permittivity tensor elements have opposite signs along orthogonal axes, provide a natural platform for such long range anisotropic guided modes to arise. Bulk hyperbolic materials support multiple types of modes in reciprocal space, including the eponymous hyperbolic dispersion region, associated with polaritons exhibiting open isofrequency contours (IFCs) and large in-plane wave vectors. However, hyperbolic materials also host non-hyperbolic yet anisotropic modes in other regions of reciprocal space, including elliptical, ghost, and lenticular modes}\cite{venturi2024visible-frequency-865, Narimanov_2017, narimanov2018dyakonov-4ec, Narimanov_2019, Ni2023}.   {When single or double interfacial effects are considered, the diversity of observed modes increases even further }\cite{Ni2023}.  {In a thin film of a type-II hyperbolic material, coupled modes similar to that of metallic thin films emerge but inherit anisotropic properties of the hyperbolic material, providing the ideal conditions to observe LRAPP behavior.} 

\begin{figure*}
    \centering
    \includegraphics[width=17cm]{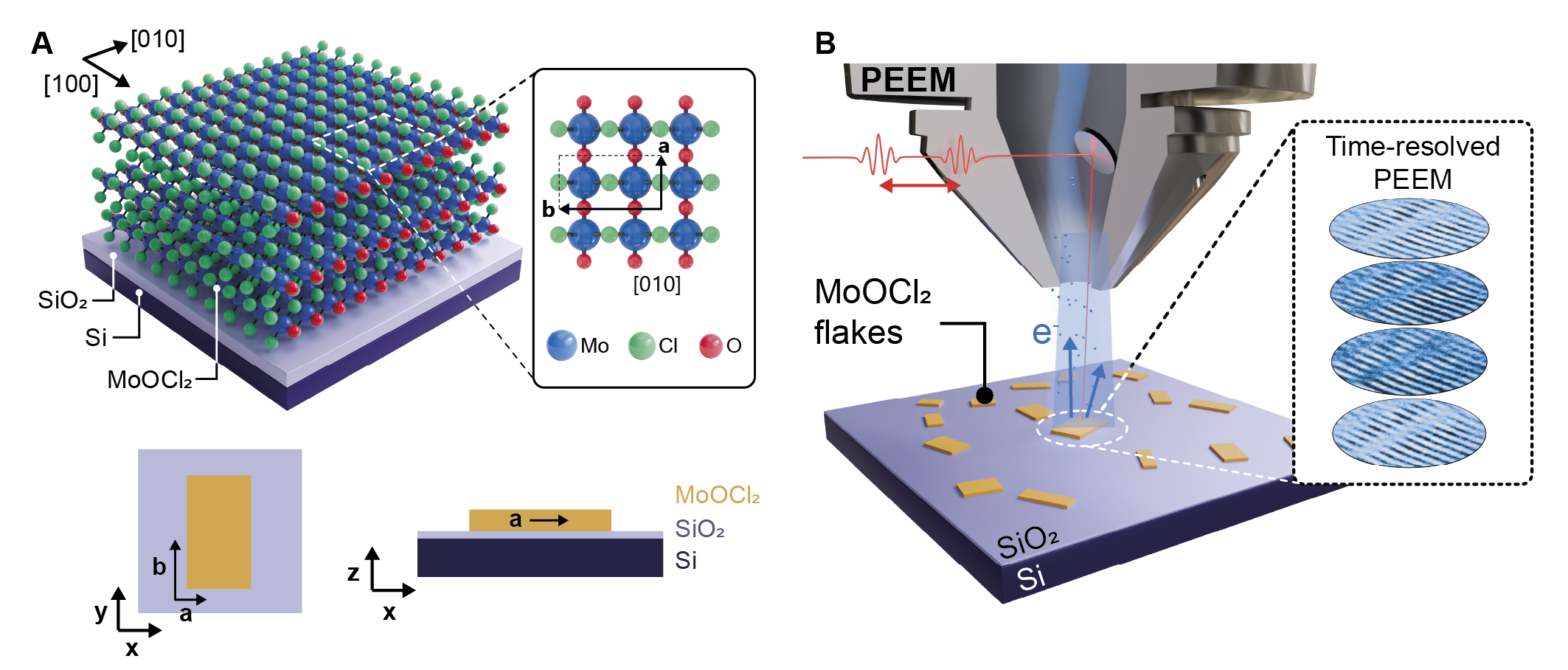}
    \caption{\textbf{Structural anisotropy and ultrafast imaging of plasmon polaritons in MoOCl$_2$.} (\textbf{A}) Crystal structure of MoOCl$_2$ and schematic of the samples used in the experiment.  {The inset shows the difference in Molybdenum bonding in the $a$ and $b$ crystal directions that gives rise to anisotropic polaritons.} (\textbf{B}) Time resolved-photoemission electron microscopy (TR-PEEM) with the delay between two interferometrically locked ultrafast laser pulses, $\Delta t$, enables direct imaging of the evolution of anisotropic plasmon polaritons in space with sub-cycle time-accuracy.}
    \label{fig:material_technique_overview}
\end{figure*}

 {Molybdenum(IV) oxydichloride} (MoOCl$_2$, Figure \ref{fig:material_technique_overview}A)  {has been identified as a hyperbolic van der Waals material that supports hyperbolic plasmon polaritons in visible and near-infrared spectral regions} \cite{venturi2024visible-frequency-865,ruta2025good-e8d,Li2025,doi:10.1021/acs.nanolett.5c03662}.  {Owing to its in-plane anisotropic dielectric and relatively low plasmonic losses, MoOCl$_2$ provides an exceptional platform for exploring long-range anisotropic plasmonic transport at technologically relevant wavelengths. Prior studies have largely investigated the hyperbolic plasmon polariton (HPP) mode in this material through reciprocal space dispersion analysis and static imaging. However, the real-space and real-time dynamics of the HPP mode and other directional plasmon polaritons remain unexplored in this material. Gaining a deeper understanding of how plasmon polaritons propagate in a hyperbolic material, interact with boundaries, and respond to different optical excitation conditions is crucial for rational device design and optimization. Direct observation of their dynamics, such as propagation speed, edge interactions, and coherence over large distances, provides critical information that extends the understanding obtained from static measurements and enables optimization of MoOCl$_2$ for low-loss, directionally selective, plasmonic and nanophotonic device applications}.

 {In this work, we provide a direct visualization of plasmon polariton dynamics in hyperbolic MoOCl$_2$ with sub-optical cycle time accuracy ($<$1 fs) using time-resolved photoemission electron microscopy (TR-PEEM, shown schematically in Figure }\ref{fig:material_technique_overview}B).  {We demonstrate a LRAPP mode in MoOCl$_2$ that has been overlooked in previous studies. This plasmonic mode exhibits propagation lengths exceeding 10 $\mu$m while maintaining the directional propagation that makes hyperbolic materials attractive for next-generation photonic devices. We directly observe the propagation of this LRAPP mode,  measure the phase and group velocities, and observe its reflection from crystal edges, all with nanoscale spatial resolution. Importantly, unlike the previously reported hyperbolic mode, the LRAPP mode we observe exhibits longer propagation lengths, faster group velocities, longer lifetimes, and intrinsically lower material losses. Our approach combines the spatial resolution of electron microscopy with the temporal resolution of ultrafast optics, enabling unprecedented insight into dynamic nanoscale light-matter interactions in hyperbolic materials.} 
\section*{Results}
Time-resolved photoemission electron microscopy (TR-PEEM) is a wide-field electron microscopy technique that leverages the high spatial resolution of electron microscopy with the time, polarization, and energy control of photon probing. As depicted in Figure \ref{fig:material_technique_overview}B, PEEM for plasmonic imaging uses an ultrafast pulsed laser to generate surface plasmon excitations, leading to electron photoemission that is magnified to produce a wide field microscope image of plasmonic fields \cite{dabrowski2020ultrafast-cc0}.  {Discontinuities on the material surface, for example a grooved coupling structure on a metal film surface} \cite{dai2020plasmonic-7e1}, {or the edge of a 2D flake as in this work, compensate for the momentum mismatch and enable the launching of plasmon polaritons without prism or nanoscale tip coupling.} PEEM can image plasmonic field distributions both with and without time-resolution. In either case, photoemission is generally a multiphoton process, and nonlinear light-plasmon interactions are required for photoemission \cite{dabrowski2016multiphoton-729, GongHess2015}.  Thus, observed fringes are the result of light-plasmon superposition. For static PEEM, the requisite superposition occurs between a plasmonic wavepacket and the same laser pulse that excited it \cite{dabrowski2016multiphoton-729, kubo2007femtosecond-d37, dai2020plasmonic-7e1}. TR-PEEM excites and probes the sample with two interferometrically locked laser pulses with a time-delay, $\Delta t$, that is controlled with a precision delay stage. Scanning the delay between two pulses enables wide-field time-resolved imaging of the plasmon dynamics of an entire spatial region simultaneously with sub-cycle temporal accuracy Figure \ref{fig:material_technique_overview}B). Unlike other approaches used to investigate plasmon dynamics in 2D materials \cite{doi:10.1126/science.aag1992}, PEEM is a wide-field imaging technique that captures the entire spatial region simultaneously. This capability simplifies high time-resolution experiments and preserves the relative phase information across the full image. Time-resolved PEEM is a well-established tool for investigating surface plasmon dynamics on noble metals \cite{dabrowski2020ultrafast-cc0, AeschlimannSteeb2007, BauerAeschlimann2007, GongHess2015}, characterizing light-matter interactions \cite{dai2020plasmonic-7e1,doi:10.1126/science.aba6415}, and has recently been extended to probe polaritons in 2D layered materials \cite{rieger2025imaging-9a2, eul2025photoemission-899, doi:10.1021/acs.nanolett.5b02363,doi:10.1021/acs.nanolett.1c03324}. Here, we exploit the nanotemporal capabilities of TR-PEEM to image  {plasmon polaritons} in MoOCl$_2$.
\subsection*{Nanoscale  {Plasmon Polariton} Imaging}
\begin{figure*}
    \centering
    \includegraphics[width=17cm]{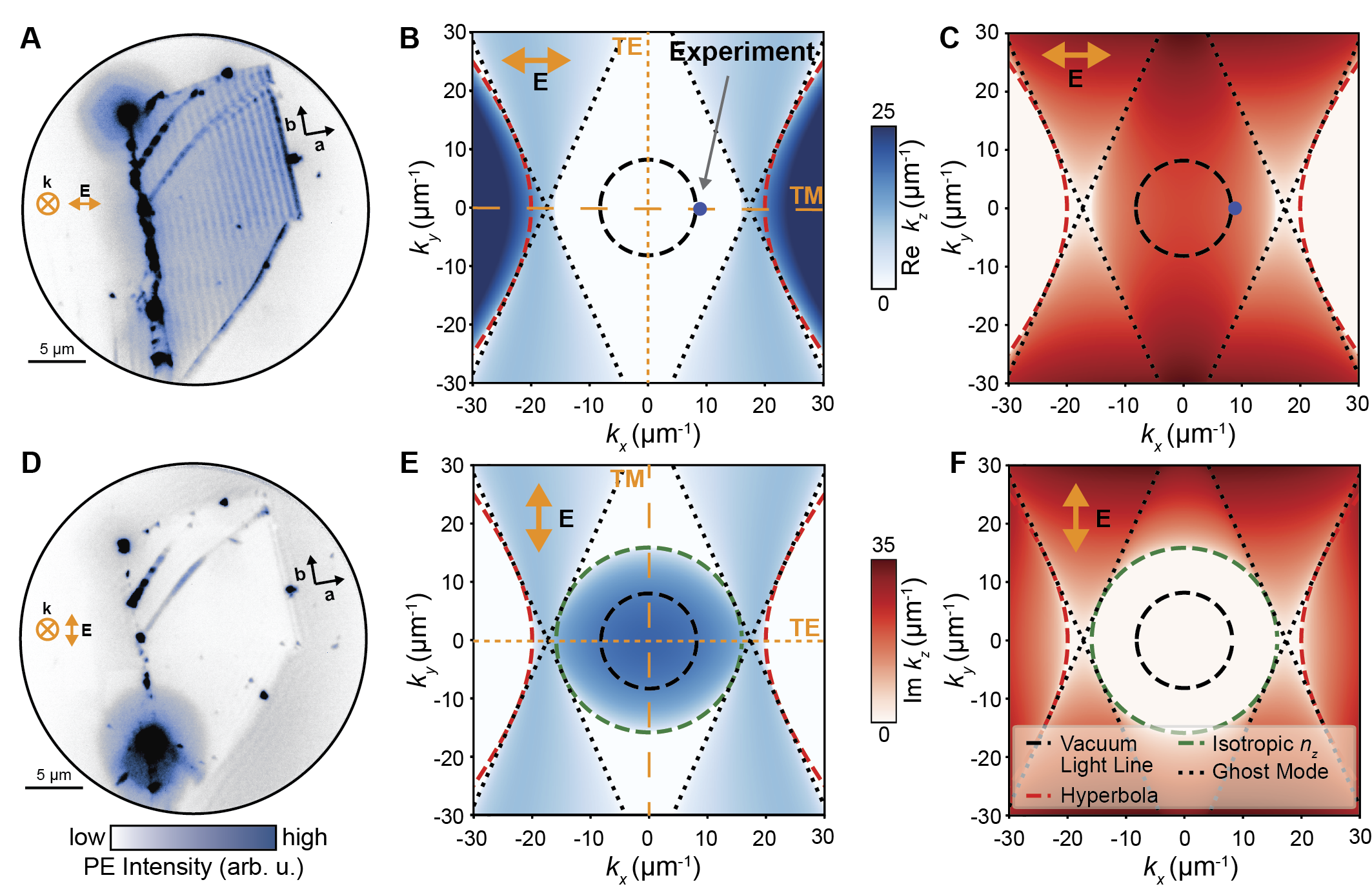}
    \caption{\textbf{Polarization-dependent anisotropic polariton dispersion in MoOCl$_2$.} {(\textbf{A}) Experimental PEEM image of MoOCl$_2$ with 1.57 eV (787 nm) photons with the laser polarized along the crystal $a$ (metallic) axis.  (\textbf{B}) and  (\textbf{C}) indicate the isofrequency contours for the real and imaginary components, respectively, of $k_z$ of MoOCl$_2$ at 1.61 eV (770 nm) for light polarized primarily along the $x$ direction. ({\textbf{D}) Experimental PEEM image of MoOCl$_2$ with 1.57 eV (787 nm) photons with the laser polarized along the crystal $b$ (non-metallic) axis.  (\textbf{E}) and  (\textbf{F}) indicate the isofrequency contours for the real and imaginary components, respectively, of $k_z$ of MoOCl$_2$ at 1.61 eV (770 nm) for light polarized primarily along the $y$ direction.}}}
    \label{fig:PD-PEEM_dispersion}
\end{figure*}
MoOCl$_2$ is a biaxial van der Waals (vdW) material with properties of a quasi one-dimensional metal with both in- and out-of-plane anisotropy in the dielectric tensor \cite{zhao2020highly-296, zhang2021orbital-selective-58f}. The primary hyperbolic region in MoOCl$_2$ is predicted to occur from the infrared to around 520 nm, where $\varepsilon'_x (\omega)< 0$ and $\varepsilon'_y (\omega)> 0$, giving rise to in-plane (type-II) hyperbolicity \cite{zhang2021orbital-selective-58f, venturi2024visible-frequency-865, Melchioni_Mancini_Nan_Efimova_Venturi_Ambrosio_2025}.  This anisotropy originates from the different character of the \ce{Mo-O} bonds along the crystalline $a$-axis ($\varepsilon_x(\omega)$) compared to the \ce{Mo-Cl} bonds along the $b$-axis  ($\varepsilon_y(\omega)$), Figure \ref{fig:material_technique_overview}A \cite{gao2021robust-7c2, zhang2021orbital-selective-58f}. Along the $b$-axis, an orbital selective Peierls phase results in \ce{Mo-Mo} dimers and flat $d_{xy}$ bands, while the $a$-axis shows metallic character with highly dispersive $d_{yz}$ and $d_{xz}$ bands \cite{zhang2021orbital-selective-58f} causing  $\varepsilon'(\omega)< 0$ along this $a$ direction and enabling plasmon propagation. This crystal-direction dependence of the plasmon propagation can be directly observed in polarization-dependent nanoscale PEEM imaging.

The plasmon polaritons measured in our PEEM experiments exhibit three hallmarks:  {directional plasmon propagation along the $a$ axis of MoOCl$_2$, strong anisotropy with respect to incident laser polarization, and dispersion of the polariton to the right of the light line.} PEEM at near-normal incidence has excellent control of light polarization in plane \cite{JoshiKing2022, SpellbergKing, ghosh2024polarization-dependent-d42, kahl2014normal-incidence-da0} enabling us to directly determine how changing the in-plane laser polarization modifies the directional plasmon propagation in MoOCl$_2$. Figures \ref{fig:PD-PEEM_dispersion}A and D show two polarization-dependent PEEM (PD-PEEM) images of a 2D MoOCl$_2$ flake excited with light polarized along the $a$ axis of the crystal or along the $b$ axis of the crystal. Interference fringes are only present along the $a$ axis; they require light polarized parallel to the $a$ axis, exciting transverse magnetic (TM) plasmons. This contrasts with isotropic materials where isotropic plasmon polaritons can be excited in arbitrary directions determined by the polarization of the incident field and the geometry of the edge that generates the momentum required for launching plasmon polaritons \cite{rieger2025imaging-9a2}. We observe similar signatures of material anisotropy in polarization-dependent optical microscopy (Movie S1) as well as polarization-resolved Raman spectroscopy (Supplementary Figure 1).

 {We can compare our PD-PEEM images with the IFCs of the two extraordinary modes present in MoOCl$_2$ in Figure} \ref{fig:PD-PEEM_dispersion}.  {In a biaxial material, there are two extraordinary modes present, which we refer to as the $a$-mode and $b$-mode due to the dominant polarization direction each exhibits (along the crystalline $a$ axis for the $a$-mode and the crystalline $b$ axis for the $b$-mode). While in general, each mode has both TE and TM polarization components, propagation along the principal axes allows for simplification. The $a$-mode is a TM mode for $k_y = 0$ (parallel to $a$ axis), but for $k_x = 0$ (parallel to $b$ axis) it is a TE mode (see Figure} \ref{fig:PD-PEEM_dispersion}B).  {For the $b$ mode the situation is reversed: for $k_y=0$, the $b$ mode is a TE mode  while for $k_x = 0$ it is a TM mode  (see Figure }\ref{fig:PD-PEEM_dispersion}E).  The polarization of each mode is illustrated in further depth in Supplementary Figure 2 { . Our observed fringes show TM polarization and propagate along the $x$ axis, and thus originate from the $a$-mode (Figures} \ref{fig:PD-PEEM_dispersion}B and C).  {This demonstrates that the experimentally measured point corresponds to a region in the IFC with a large Im$(k_z)$,  meaning that the observed mode is a surface wave evanescent in the $z$ direction, distinct from both the ghost modes (with mixed polarization) and the HPP modes (dominated by Re($k_z$)). Both the ghost modes and the HPP propagate in the bulk, similar to waveguide modes.} 

\begin{figure*}
    \centering
    \includegraphics[width=17cm]{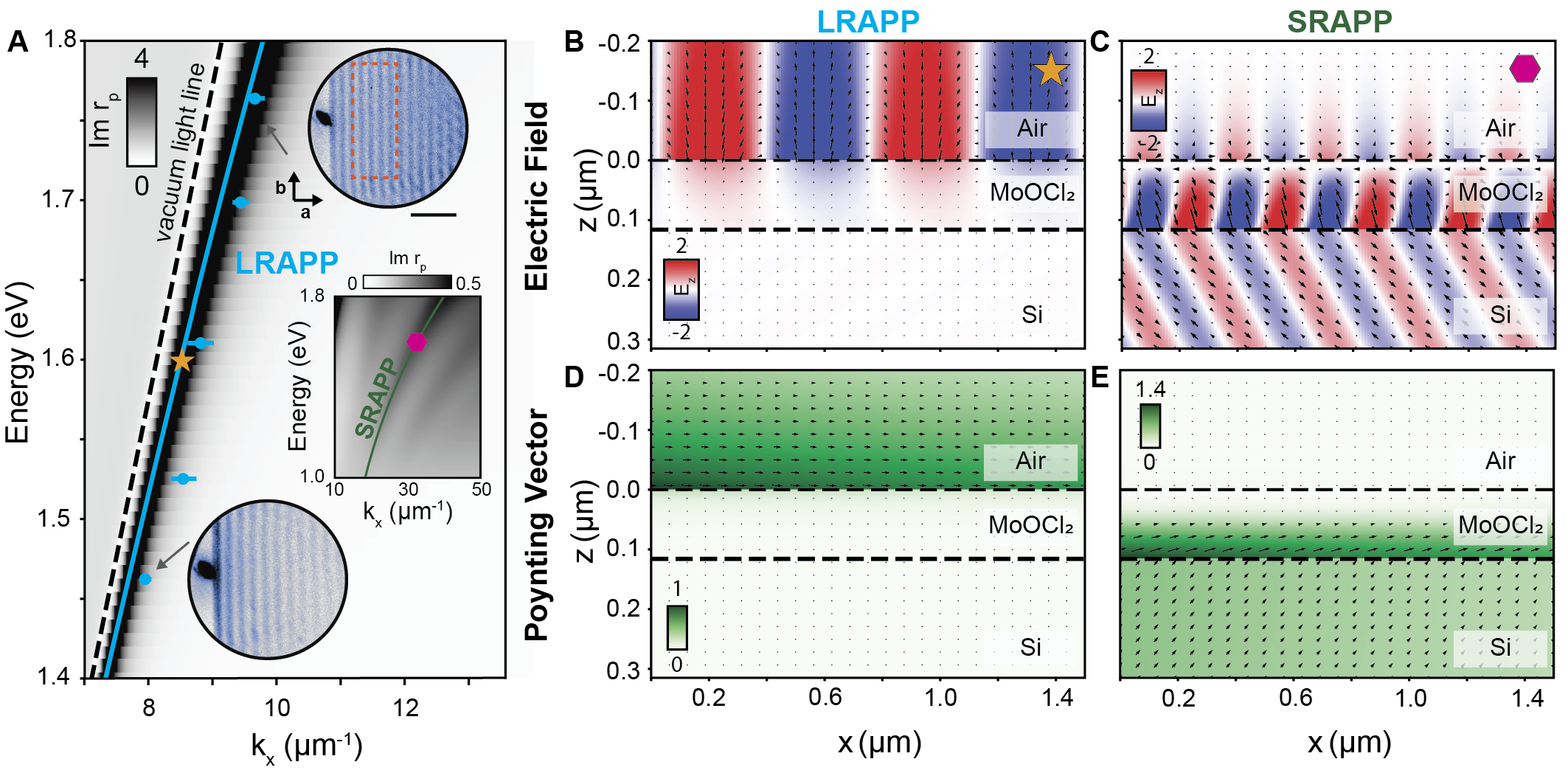}
    \caption{\textbf{Long- and short-range anisotropic plasmon polaritons in MoOCl$_2$.} (\textbf{A}) Experimental measurements of plasmon polaritons in MoOCl$_2$ extracted from static PEEM measurements (insets, scale bar 3 $\mu$m) compared with a model of the imaginary component of the pole of the reflectivity coefficient and analytical models for both the long-range and short-range anisotropic plasmon polaritons (LRAPP and SRAPP). Calculations of the Electric Field (\textbf{B}) and Poynting vector (\textbf{C}) for the LRAPP and the Electric Field (\textbf{D}) and Poynting vector (\textbf{E}) for the SRAPP for the energies shown in (\textbf{A}). Also see Supplementary Figure 6 for charge density for these modes.}
    \label{fig:dispersion}
\end{figure*}

Repeating static PEEM experiments as a function of photon energy, we observe that the fringe spacing changes with energy (Figure \ref{fig:dispersion}A), based on which we extract the dispersion of the plasmon polariton (Figure \ref{fig:dispersion}A and Figure Supplementary 3). The experimental data points are in good agreement with analytical calculations of the dispersion along the $a$ axis from reported dielectric functions \cite{Melchioni_Mancini_Nan_Efimova_Venturi_Ambrosio_2025},  { as well as the polarization specific $r_{pp}(k_x,k_y)$ reflection coefficient isofrequency plot for an infinitely thick MoOCl$_2$ flake in Supplementary Figure 4}. Notably, our experiments also  {corroborate polarization-dependent transmission and reflectivity measurements} \cite{Melchioni_Mancini_Nan_Efimova_Venturi_Ambrosio_2025} that the energy region over which MoOCl$_2$ is hyperbolic is narrower than predicted from theoretical models \cite{zhao2020highly-296}. We do not see evidence of plasmon polaritons in experiments at 2.4 eV (516 nm) and higher energies (Supplementary Figure 5) as theoretically predicted by \citet{zhao2020highly-296}  {but consistent with the experimentally reported dielectric function by} \citet{Melchioni_Mancini_Nan_Efimova_Venturi_Ambrosio_2025}. 

\begin{figure*}
    \centering
    \includegraphics[width=17cm]{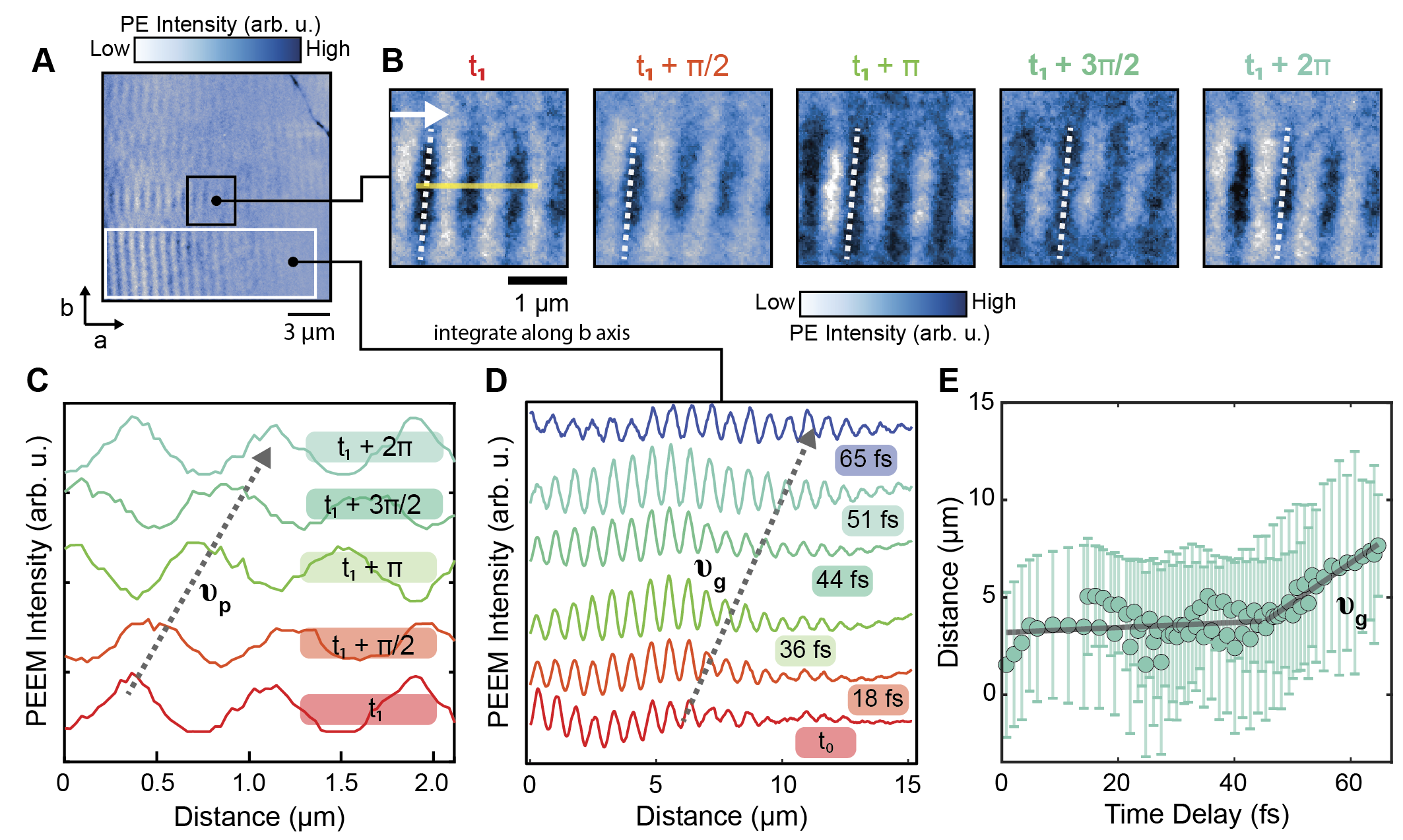}
    \caption{\textbf{Sub-cycle interferometric tracking of anisotropic plasmon polariton propagation.} (\textbf{A}) Image at $\Delta t= 0$ for an interferometric PEEM experiment with h$\nu=1.60$ eV. (\textbf{B}) Expansion of the black box from (\textbf{A}) for different phase delays of the two phase-locked laser pulses. For this photon energy a phase shift of $\pi/2=0.645$ fs. The same peak is traced in a white dotted line through each image. (\textbf{C}) Linecuts along the yellow line in (\textbf{B}) for each phase delay. (\textbf{D}) Linecuts from the white box in (\textbf{A}) where the intensity has been integrated along the $b$ axis showing the group velocity of the  {LRAPP} wavepacket as it propogates from the edge. (\textbf{E}) The fit gaussian center of the  {LRAPP} wavepacket as it propagates away from the edge as a function of time. The group velocity is fit to the propagating wave between 44 and 64 fs. $v_p$ and $v_g$ indicate phase and group velocitues, respectively.}
    \label{fig:group_phase_velocity}
\end{figure*}

In addition to comparisons of our experimental dispersion to the imaginary component of the pole of the reflectivity coefficient for the  {vacuum/MoOCl$_2$/ 2 nm SiO$_2$/Si} sample geometry, we also compared our results to calculations based on a transfer matrix method (as described in the Supplementary Information) to extract the electric field and Poynting vector distributions for different modes,   {shown in Figures \ref{fig:dispersion}B-E}. The MoOCl$_2$ flakes investigated in this and other works are typically less than 200 nm thick\cite{ruta2025good-e8d, doi:10.1021/acs.nanolett.5c03662, Li2025}.  Such thin plasmonic layers cause a coupling of plasmonic modes resulting in what is referred to as short-range and long-range plasmon polaritons\cite{Großmann2021, doi:10.1126/sciadv.1700721}. Our calculations reveal a similar effect in MoOCl$_2$ with the added feature of in-plane anisotropy. We identify these modes as a long-range anisotropic plasmon polariton (LRAPP), as described earlier, and a short-range anisotropic plasmon polaritons (SRAPP) as indicated in Figure \ref{fig:dispersion}A (also see  {Supplementary Figure 6 for charge density distribution of these two plasmon modes and} Supplementary Figure 7  {for thickness dependence of the LRAPP mode). Notably the SRAPP mode corresponds to the mode previously explored as the HPP mode in MoOCl$_2$ as this mode occurs in the unbounded portion of the IFCs.  The field distributions and Poynting vectors, calculated based on the transfer matrix method \cite{passler2017generalized-a63, raab2024surface-28e, rieger2025imaging-9a2} as discussed in Supplementary Note 3  for the LRAPP and SRAPP are shown in Figures \ref{fig:dispersion}B-E,  demonstrating the characteristic pinwheel shape for plasmon polaritons and Poynting vectors localized on the upper and lower interfaces. Similar kinds of polaritonic modes, but associated with phonons, have been reported in the THz portion of the electromagnetic spectrum in calcite where they are referred to as directional leaky polaritons at anisotropic crystal interfaces \cite{Ni2023}.  {Using that nomenclature, the LRAPP discussed here is a directional leaky plasmon polariton which we term as a long-range anisotropic plasmon polariton in order to distinguish from the other anisotropic plasmon polariton mode, the SRAPP.}

Related so-called ghost modes \cite{ma2021ghost-2bb},  which show complex \textbf{k$_z$} even without intrinsic losses, have been reported in MoOCl$_2$ \cite{venturi2024visible-frequency-865}. However, we rule these modes out based on the surface sensitivity of our imaging method and the long propagation lengths we observe. Ghost modes arise at energies where $\min{(\varepsilon_x, \varepsilon_y)} < \varepsilon_z <  \max{(\varepsilon_x, \varepsilon_y})$, occur within the bulk, and exhibit significant losses \cite{venturi2024visible-frequency-865, narimanov2018dyakonov-4ec}. In contrast, PEEM is a highly surface sensitive technique, relying on photoemission for signal detection \cite{SeahDench1979}, and the fields we observe propagate over distances exceeding $10\ \mu$m (discussed later in this paper), which is inconsistent with the short decay length one would expect for lossy ghost modes. Overall, based on the directional plasmon propagation, dispersion,  {strong sensitivity to laser polarization, and eliminating possible contributions from ghost modes, we assign the plasmon polaritons observed in our experiments as LRAPPs in MoOCl$_2$}.

\subsection*{Space-Time Imaging of  {Plasmon Polaritons}}
Time-resolved imaging of the dynamics of LRAPPs in space and time allows for direct extraction of their phase and group velocities and observation of the interaction of LRAPPs with nanoscale structures such as flake edges. In a TR-PEEM experiment, a Michelson interferometer (Supplementary Figure 8) creates a pair of phase-locked laser pulses that are controlled in time with a mechanical stage which can precisely create time-steps $<1$ fs, to directly image the dynamics of  {LRAPPs}. The two laser pulses incident on the sample create a complex set of interferences due to light-light, light-plasmon, and plasmon-plasmon interactions. To isolate the dynamic contribution, we perform pixel wise time-domain Fourier filtering of the raw data \cite{dai2020plasmonic-7e1,spektor2019mixing-79c,dabrowski2020ultrafast-cc0}, as shown in Supplementary Figure 9, and extract the in-plane light-plasmon interactions that are first order in the driving laser frequency.  {Please note that our laser pulse duration is approximately 30 fs, therefore, we primarily focused our analysis, including the PEEM snapshots in Figure }\ref{fig:group_phase_velocity}  {and the calculations of phase and group velocities, to time delays at least 30 fs after time-zero (pump-probe temporal overlap). This is because light-light interactions dominate near time-zero, whereas light-plasmon interferences are dominant at longer delays, as established in previous interferometric TR-PEEM studies }\cite{spektor2019mixing-79c, davis2017subfemtosecond-8f9, dai2020plasmonic-7e1}.

These dynamics enable us to extract the phase ($\upsilon_\text{p}$) and group ($\upsilon_\text{g}$) velocities of  {LRAPPs} in MoOCl$_2$, directly observing the impact of MoOCl$_2$'s dispersion on the dynamics of  {LRAPP} propagation. A snapshot of an interferometric TR-PEEM experiment with 1.60 eV photons is shown in Figure \ref{fig:group_phase_velocity}A. Zooming in on a portion of the flake, Figure \ref{fig:group_phase_velocity}B, one can directly observe the dynamics of the HPPs within an optical cycle (\textasciitilde 2.6 fs). The data clearly shows oscillatory behavior as the  {LRAPP} interacts and interferes with the laser pulse. While the carrier envelope phase (CEP) of the driving laser is not stabilized between different repetitions of the laser, the relative CEP between two pulses within one repetition is constant and has a locked phase relationship. This provides access to directly image the sub-cycle dynamics and extract the phase velocity of the  {LRAPP}. Line profiles from the yellow line in Figure \ref{fig:group_phase_velocity}B are shown in Figure \ref{fig:group_phase_velocity}C. Tracing the spatial dynamics of a peak in the line profile as a function of time, we extract a plasmon phase velocity of 2.93 $\pm$ 0.03 $\times \ 10^8$ m/s, or 0.98 $\pm$ 0.01 $c$, in good agreement with theoretical calculations as described in Supplementary Figures 10 and .
\begin{figure*}
    \centering
    \includegraphics[width= 17cm]{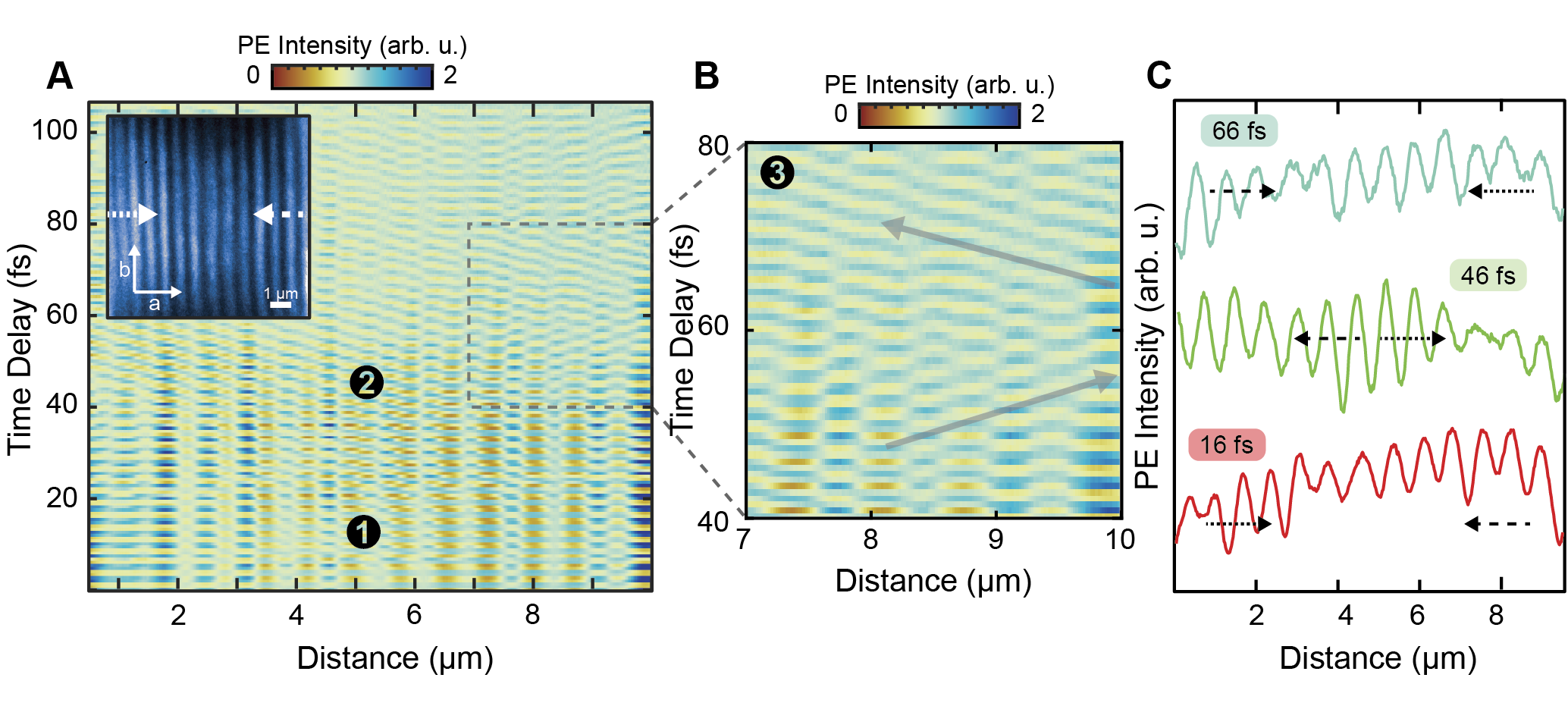}
    \caption{\textbf{Spatiotemporal dynamics and edge reflection of anisotropic plasmon polariton wavepackets.} (\textbf{A}) Space-time contour plot of a PEEM movie with 1.60 eV (Movie S2 on Flake C) where  {LRAPPs} are launched from the edges of the inset flake snapshot along the $a$ axis. As a function of time the  {LRAPP} wavepackets propagate away from the edges (1), pass each other in the middle of the flake (2), and reflection off of the other edge (3). A zoom in of the reflection dynamics are shown in (\textbf{B}) where the arrows trace the dynamics of the  {LRAPP} wavepacket. (\textbf{C}) Accompanying line cuts integrated along the $b$ axis where the dotted and dashed arrows demonstrate the direction of the  {LRAPPs} launched from the left and right edges, respectively as they undergo propagation.}
    \label{fig:tr_reflection}
\end{figure*}

From the same experiment, one can also directly visualize how a plasmon wavepacket moves away from the edge of a MoOCl$_2$ flake and determine the corresponding group velocity. As shown in Figure \ref{fig:group_phase_velocity}D, the initial excitation fringe profile moves to the right as a function of time. Fitting these fringe profiles to a Gaussian wavepacket envelope (description in the Supplementary Information and Supplementary Figures 12-14), we can quantitatively measure how the plasmon wavepacket moves in time (Figure \ref{fig:group_phase_velocity}E). There are two regions in the dynamics, a quasi-stationary initial period, and then a propagating period. Such quasi-stationary plasmon polaritons have been observed previously in experiments on noble metal surfaces due to the plasmon interaction with the same pulse that excited it \cite{PodbielHeringdorf2017}. In the propagating period we extract a plasmon group velocity of 2.0 $\pm$ 0.3 $\times \ 10^8$ m/s, or 0.68 $\pm$ 0.09 $c$.  This experimental group velocity is slower than calculations based on an analytical model of the  {LRAPP} dispersion, where $\upsilon_\text{g}=\partial \omega(k)/\partial k$ in Supplementary Figure 10 predicts a plasmon group velocity of 0.84 $c$.  {One possible reason for this discrepancy could be that the analytical model used to calculate $\upsilon_\text{g}$ does not account for loss mechanisms due to electron-electron scattering, electron-phonon scattering and scattering from surface defects, which can reduce the plasmon group velocity} \cite{Ghorashi_highly_2024}.

\subsection*{Plasmon Polariton Propagation Lengths and Edge Interactions}
The supplementary movies (Movie S2 and S3) of the time-resolved  {LRAPP} dynamics in a MoOCl$_2$ flake showcase two notable findings, propagation lengths greater than 10 $\mu$m, and the ability to image the dynamic interaction of  {LRAPPs} with flake edges. Movie S2 shows the launching of  {LRAPPs} from two edges of a MoOCl$_2$ flake (white arrows indicated in the inset of Figure \ref{fig:tr_reflection}A). These two plasmonic wavepackets propagate away from the flake edge, pass each other in the middle of the flake, creating standing wave dynamics,  reflect off of the opposite side, and continue to propagate. Figure \ref{fig:tr_reflection}A shows a space-time contour plot of the flake from this movie, integrated in the $b$ direction, showing the photoemission intensity as a function of space and time. The initial excitation of the  {LRAPP} wavepackets occurs at point (1), the standing wave dynamics at point (2) and the zoomed in region of Figure \ref{fig:tr_reflection}B shows the reflection of the  {LRAPPs} at (3), highlighted by the gray arrows. The propagation direction of  {LRAPPs} at each of these stages is further evident from the intensity profile of Figure \ref{fig:tr_reflection}C. As can be observed from Movie S2 (also see Associated Content section and Movie S3), these dynamics continue as the wavepacket propagates back towards the original edge and reflects again. 

Based on the plasmon dynamics shown in Movie S2 across the flake (Figure \ref{fig:tr_reflection}A) which is approximately 11 $\mu m$ wide, the  {LRAPPs} traverse the flake and reflect twice at the edges, corresponding to a total travel distance exceeding 33 $\mu m$. This indicates that  {LRAPP} propagation length is greater than 10 $\mu$m. This is similar to other estimates based on the group velocity and lifetime of  {LRAPPs} in MoOCl$_2$ described in the Supplementary Information. Such propagation lengths suggest that MoOCl$_2$ is a promising material for development in nanophotonic devices \cite{liu2016fundamental-abc}.

Our demonstrated ability to directly image plasmon polariton reflections from a MoOCl$_2$ flake edge with TR-PEEM also opens numerous further avenues for investigation. One of the unusual properties of  {LRAPPs} are the ways they may interact with material discontinuities, leading to effects like negative refraction at interfaces with other biaxial materials \cite{high2015visible-frequency-498, Zhang_Zheng_Chen_Qiu_2022, Sternbach_Moore_Rikhter_Zhang_Jing_Shao_Kim_Xu_Liu_Edgar_et_al_2023, Liu_Huang_2023} or unconventional Goos-H\"{a}nchen reflection phase shifts  \cite{goos1947ein-bb6, melentiev2021spp-fa1, kang2017goos-hanchen-7aa, yallapragada2016observation-20e}. While nanopatterning of MoOCl$_2$ is likely required for such experiments, our results demonstrate that MoOCl$_2$ and TR-PEEM would be a fruitful playground to realize such experiments, testing fundamental questions in physics and nanophotonics.

\section*{Discussion}
We provide a direct spatiotemporal visualization of anisotropic plasmon polaritons in the visible portion of the electromagnetic spectrum in a natural material and identify a previously overlooked plasmon polariton, the long range anisotropic plasmon polariton, with much longer propagation lengths than the known hyperbolic plasmon polaritons. The performance metrics we have measured, including propagation lengths exceeding 10 $\mu$m, near-light-speed velocities, and robust directional propagation, position MoOCl$_2$ as a compelling alternative to conventional plasmonic materials for visible-light applications. Such long polaritonic propagation lengths are a critical requirement for practical integration into optoelectronic devices and can provide advantages such as high energy efficiency, reliable signal transfer between components and the potential for scalable on-chip architectures }\cite{Gramotnev_Bozhevolnyi_2010}.  {Notably though, this long-range anisotropic mode coexists with the hyperbolic modes, which have greater confinement, suggesting the possibility of exploiting both modes or exchanging energy or momentum between the two modes in future applications.

Real-time visualization of  {LRAPP} dynamics provides device designers with powerful tools to predict performance, optimize geometries, and troubleshoot fabrication challenges, {an essential advancement for enabling rational device design and fabrication}. Our demonstration of  {LRAPP} reflection dynamics and edge interactions offers insights for designing waveguides, couplers, and other essential components of integrated photonic circuits. This capability also enables direct experimental testing of theoretical predictions including negative refraction \cite{Argyropoulos:13,Sternbach_Moore_Rikhter_Zhang_Jing_Shao_Kim_Xu_Liu_Edgar_et_al_2023}, Goos-H\"{a}nchen shifts \cite{Sobucki2025}, and other unusual phenomena in hyperbolic media \cite{Zhou:19}. Beyond applications, our technique opens pathways to investigate fundamental questions about the limits of light confinement, the role of material anisotropy in electromagnetic propagation, and the interplay between electronic correlations and collective excitations in van der Waals materials. Access to dynamic information at ultrafast timescales promises to uncover physics that static measurements alone cannot reveal.

Looking forward, this work establishes visible-frequency  {directional} plasmonics as a platform for next-generation nanophotonic technologies \cite{lee2022hyperbolic-c0d,Chen2025}. While  {anisotropic} plasmonic responses are often engineered through complex metamaterial architectures \cite{https://doi.org/10.1002/lpor.201400457} requiring complicated fabrication process, \ce{MoOCl_2} hosts visible-range  {anisotropic plasmon polaritons} that can be accessed via simple mechanical exfoliation. The combination of  {long-range polaritons with low loss}, enhanced light-matter interactions, air-stability, and room-temperature operation positions MoOCl$_2$ and related materials at the forefront of applications ranging from quantum information processing to ultra-sensitive biosensing \cite{wang2024planar-76d}. More broadly, our spatiotemporal imaging approach provides a distinct experimental paradigm for studying collective excitations in anisotropic materials, promising to accelerate discoveries across condensed matter physics, nanophotonics, and quantum optics \cite{grankin2023interplay-fe4,Aigner2022,Gomez-Diaz:15}.
\section*{Methods}
 \subsection{Sample preparation and characterization}
Single crystals of MoOCl$_2$ were purchased from hq graphene. Using mechanical exfoliation with scotch tape and transfer with PDMS viscoelastic stamping, flakes of MoOCl$_2$ were deposited onto silicon wafers (n-doped Si(100), $\rho=1-10\,\mathrm{\Omega}$) with  {2 nm  \ce{SiO_2} layer on top,} purchased from University Wafer. This process produces separated flakes with lateral sizes of tens of microns (see Supplementary Figure 15A). All exfoliation was done in a $N_2$ glove box prior to transfer into ultra-high vacuum (UHV) to maintain high quality samples even though no evidence for atmospheric degradation was observed. Atomic force microscopy was conducted in the tapping mode to determine the thickness of the flakes (see Supplementary Figure 15C). Raman spectroscopy (Supplementary Figure 1) was carried out in a HORIBA LabRAM HR Evolution confocal Raman microscope.  {The Raman spectra shown in Supplementary Figure 1 are consistent with other reports in the literature for MoOCl$_2$ \cite{Melchioni_Anisotropic_2026, Minnekhanov_hyperbolic_2025}.}

\subsection{PEEM experiments}
All PEEM experiments were performed under ultra-high vacuum ($<1 \times 10^{-10}\,\mathrm{mbar}$) using a photoemission electron microscope from Focus GmbH. The microscope has a spatial resolution better than $40$ nm. One-photon photoemission microscopy (1P-PEEM) was carried out with a $100\,\mathrm{W}$ broadband mercury arc lamp ($h\nu \le 5.1$ eV) under grazing incidence ($65\,\mathrm{^{\circ}}$ to the surface normal) to locate the MoOCl$_2$ flakes on the surface and obtain an overview of the flake geometry (Figure 15B).

Initially unoccupied intermediate states, such as plasmon resonances, may be pumped and probed in nP-PEEM (with $n \ge 2$) so that they may contribute to the PEEM signal. Monochromatic nP-PEEM ($n=4$) experiments were conducted with fundamental ($1.35 \le \mathrm{1h \nu} \le 1.9$ eV, $5-30\,\mathrm{nJ}$ pulse energy, $30-40\,\mathrm{fs}$ pulse duration) of a home-built optical parametric amplifier (OPA). The OPA was pumped by the second harmonic ($2 \mathrm{h \nu} = 2.4$ eV) of a Coherent Monaco with a repetition rate of $4\,\mathrm{MHz}$. The laser spot sizes on the sample were approximately $300$ to $1000$ $\mu m^2$. For all time resolved results presented in this manuscript with excitation $h\nu = 1.6$ eV, photoemission from MoOCl$_2$ is a 4PPE process, as shown in Supplementary Figure 16.

The linearly polarized laser beam was directed to the sample either at near-normal incidence ($4\,\mathrm{^{\circ}}$ to the surface normal) via a rhodium mirror inside the microscope column (see Figure 1B). In this configuration, the laser polarization is effectively in the plane of the sample surface. $\mathrm{\lambda/2}$ waveplates were used to tune the in-plane laser polarization with respect to the edges of the MoOCl$_2$ flakes. Details of laser polarization calibration can be found in references \citet{JoshiKing2022, SpellbergKing} and in the Supplementary Figure 17.

Interferometric time-resolved experiments were performed with 1.60 eV (774 nm) photons with controlled time-delay and phase with a Michelson Interferometer (Supplementary Figure 8). Photons are incident on a 50/50 beam splitter (Thorlabs BSW26R), half one path is directed onto a 0 degree mirror while the other is directed onto a 0 degree mirror mounted to a mechanical delay stage (motorized linear stage, Newport, XMS50-S). The pulses then are recombined with the same beamsplitter and directed onto a series of steering mirrors before being focused into the UHV chamber through a lens and Rhodium mirror mounted in the microscope column.

\subsection{Simulations of HPPs}
Numerical simulations were used to model the hyperbolic polaritions. For these simulations we consider a stack consisting of a MoOCl$_2$ flake with thickness $d$ on top of $2\,\mathrm{nm}$ native SiO$_2$ on top of $380\,\mathrm{\mu m}$ silicon(100). The dielectric functions of the SiO$_2$ and silicon were obtained from the SOPRA data bank (\url{https://www.sspectra.com/sopra.html}) and the Handbook of Optical Constants of Solids, respectively \cite{borghesi1991handbook-dde}. We conducted our simulations using the  {experimentally determined dielectric function of MoOCl$_2$ from \citet{Melchioni_Mancini_Nan_Efimova_Venturi_Ambrosio_2025} in the $x$ and $y$ directions and the calculated DFT dielectric function of MoOCl$_2$ from \citet{zhao2020highly-296} for the dielectric function in the $z$ direction as there is no experimental data for this direction.}

A transfer matrix method \cite{passler2017generalized-a63, passler2023layer-resolved-849} was used to calculate the dispersion relationships for MoOCl$_2$. This approach has been successfully employed before to model polaritons observed with scattering near-field optical microscopy \cite{dai2014tunable-616, ruta2025good-e8d, mooshammer2022in-plane-6c3} as well as PEEM \cite{rieger2025imaging-9a2, raab2024surface-28e}. Details and derivations can be found in the Supplementary Information and references \citet{rieger2025imaging-9a2, raab2024surface-28e} as well as  in the associated code for this manuscript on GitHub.

\section*{Data Availability}
The static and time-resolved PEEM data generated in this study have been deposited in the Zenodo database under accession code https://doi.org/10.5281/zenodo.15685600 \cite{zenodo}. The sample characterization data used in this study are available in the Supplementary Information.

\section*{Code Availability}
The Python code and Jupyter Notebooks used for calculations in this publication are available on GitHub under the accession code https://doi.org/10.5281/zenodo.18684526 \cite{sbking_pchem2026}.

\section*{References}
\bibliography{references_formatted}% Produces the bibliography via BibTeX.

\section*{Associated Content}
The Supplementary Information includes  {polarized Raman spectroscopy, details of the polarization of modes of MoOCl$_2$, calculations of the dispersion, the reflection coefficients, and the field distributions for the various modes, static PEEM images, charge density plots, thickness dependence of the different modes, a schematic of the interferometer, details of Fourier filtering, the phase and group velocity measurements, characterization of MoOCl$_2$ flakes, the power dependence of the PEEM images, calibration of the laser polarization, and calculations of the propagation lengths.} Supplementary movie S1 shows polarization dependent optical microscopy of Flake B emphasizing the strong optical anisotropy in \ce{MoOCl_2}. Supplementary movies S2 and S3 show dynamics for two different spatial regions for Flake C demonstrating space time propagation of HPPs. Movie S2 presents full spatiotemporal evolution of HPPs as they launch from edges, pass one another, and reflect back, as discussed in Figure \ref{fig:tr_reflection}. This particular region of the flake features straight, uniform edges, which support well-defined counter-propagating HPPs. In contrast, Movie S3 depicts  {LRAPP} propagation from a different and wider region of Flake C, where the surface and edges are less uniform. Consequently, the  {LRAPPs} in this region propagate in various direction across the areas of the flake, depending on their respective point of launch.

\begin{acknowledgments}
We acknowledge Ryo Mizuta Graphics for providing 3D illustrations as part of this manuscript's figures and Dr. Karen Watters for help with the scientific writing and editing.

This work was funded by the Air Force Office of Scientific Research (FA9550-22-1-0224). This work made use of the shared facilities at the University of Chicago Materials Research Science and Engineering Center, supported by the National Science Foundation under award number DMR-2011854. This work was completed in part with resources provided by the University of Chicago’s Research Computing Center. A.G. acknowledges support from a MRSEC-funded Kadanoff-Rice fellowship (DMR-2011854).
\end{acknowledgments}

\section*{Author Contributions Statement}
S.B.K. conceived of the study. A.G. and C. R. performed the experiments with help from J.L.S. and M.M.. A.G. and C.R. performed the simulations. A.M. provided the samples. J.R. built the interferometer. A.G., C.R., and J.L.S. performed the data analysis with help and input from S.B.K.. S.B.K., A.G., C.R., and J.L.S. prepared the figures. A.G., C.R. and S.B.K. wrote the manuscript with contributions from all authors.  S.B.K. supervised the project and procured funding.
\section*{Competing Interests Statement}
The authors declare no competing interests.

\end{document}